\documentclass[%
 aip,
 amsmath,amssymb,
 reprint,%
]{revtex4-1}

\usepackage[dvipdfmx]{graphicx}
\usepackage{dcolumn}
\usepackage{bm}

\usepackage[utf8]{inputenc}
\usepackage[T1]{fontenc}
\usepackage{mathptmx}

\begin{document}

\preprint{AIP/123-QED}

\title{A measurement method for responsivity of microwave kinetic inductance detector by changing power of readout microwaves}

\author{H. Kutsuma}
\altaffiliation{h.kutsuma@astr.tohoku.ac.jp}
\affiliation{Astronomical institute, Tohoku University, 6-3 Aramaki, Aoba-ku, Sendai 980-8578, Japan}
\affiliation{RIKEN Center for Advanced Photonics, 2-1 Hirosawa, Wako 351-0198, Japan}
\author{M. Hattori}
\affiliation{Astronomical institute, Tohoku University, 6-3 Aramaki, Aoba-ku, Sendai 980-8578, Japan}
\author{R. Koyano}
\affiliation{Saitama University, Shimo-Okubo 255, Sakura-ku, Saitama 338-8570, Japan}
\author{S. Mima}
\author{S. Oguri}
\author{C. Otani}
\affiliation{RIKEN Center for Advanced Photonics, 2-1 Hirosawa, Wako 351-0198, Japan}
\author{T. Taino}
\affiliation{Saitama University, Shimo-Okubo 255, Sakura-ku, Saitama 338-8570, Japan}
\author{O. Tajima}
\affiliation{Kyoto University, Kitashirakawa-Oiwakecho, Sakyo-ku, Kyoto 606-8502, Japan}

\date{\today}

\begin{abstract}
Superconducting detectors are a modern technology applied in various fields. The microwave kinetic inductance detector (MKID) is one of cutting-edge superconducting detector. It is based on the principle of a superconducting resonator circuit. A radiation entering the MKID breaks the Cooper pairs in the superconducting resonator, and the intensity of the radiation is detected as a variation of the resonant condition. Therefore, calibration of the detector responsivity, i.e., the variation of the resonant phase with respect to the number of Cooper-pair breaks (quasiparticles), is important. We propose a method for responsivity calibration. Microwaves used for the detector readout locally raise the temperature in each resonator, which increases the number of quasiparticles. Since the magnitude of the temperature rise depends on the power of readout microwaves, the number of quasiparticles also depends on the power of microwaves. By changing the power of the readout microwaves, we simultaneously measure the phase difference and lifetime of quasiparticles. We calculate the number of quasiparticles from the measured lifetime and by using a theoretical formula. This measurement yields a relation between the phase response as a function of the number of quasiparticles. We demonstrate this responsivity calibration using the MKID maintained at 285$\,$mK. We also confirm consistency between the results obtained using this method and conventional calibration methods in terms of the accuracy.
\end{abstract}

\maketitle

Superconducting detectors have greater sensitivity compared with conventional technologies. This is because the gap energies of superconducting devices are lower than those of semiconductor devices by three orders of magnitude. Therefore, this technology is applied in various scientific fields. The microwave kinetic inductance detector (MKID) [\onlinecite{MKIDs}] is one of cutting-edge superconducting detector. It is based on the principle of superconductive resonator circuits, i.e., each detector consists of a simple inductor and a capacitor. These circuits are implemented in  superconducting films on a substrate made of silicon or sapphire. The resonant frequency of each detector is defined by the geometry of the film. MKIDs naturally provide a multiplex readout in the frequency domain using microwaves. Therefore, compared with other types of superconducting detectors, there is an advantage in constructing a large format array. In particular, the field of  radio astronomy has started using antenna-coupled MKID arrays [\onlinecite{DESHIMA, NIKA, GB_nagasaki}]. 

The detection mechanism of the MKIDs is simple. A radiation entering the detector breaks the Cooper pairs in the superconducting film, and changes the inductance of the resonator. It also changes the resonant condition of microwaves coupled to the inductors via a feedline. We measured the  intensities of  each detector signal based on the variations of each resonance. We used the resonant phase ($\theta$) for the measurements in this study. Broken Cooper pairs are called quasiparticles. The mean time for the recombination of the broken Cooper pairs is called the quasiparticle lifetime ($\tau_\mathrm{qp}$). The number of quasiparticles ($N_\mathrm{qp}$) was calculated with material properties, geometries, and the $\tau_\mathrm{qp}$ as described in  [\onlinecite{tauqp_Nqp}, \onlinecite{pieter2}]. 

In real applications, we have to calibrate the responsivity of the detector, i.e., $\mathrm{d}\theta/\mathrm{d}N_\mathrm{qp}$. It is one of important indicators to evaluate the sensitivity of MKID, and it is difficult to develope the MKID without this information. There are two conventional methods: a calibration by changing the physical temperature of a MKID device, and another calibration through a power spectral density. This paper proposes the third caliblation method for the responsivity, i.e., simultaneous measurements of $\theta$ and $\tau_\mathrm{qp}$ through the power change of readout microwaves.

Changing the physical temperature of a MKID device is the most popular calibration method [\onlinecite{MKIDs},  \onlinecite{gao2008}]. We increase $N_\mathrm{qp}$ by warming up the device by using a heater set in the cryostat. The relation between the temperature $(T)$ and $N_\mathrm{qp}$ is given by the following formula [\onlinecite{BCS}]: 
\begin{equation}
\label{eq:nqp}
N_\mathrm{qp} = 2N_0V\sqrt{2\pi k_\mathrm{B}T\Delta}\exp{\left(-\frac{\Delta}{k_\mathrm{B}T}\right)},
\end{equation}
where $N_0$ is the single spin density of the states at the Fermi energy (e.g., $N_0 \sim 1.74\times 10^{10}\,\mathrm{eV}^{-1}/\mu\mathrm{m}^3$ for an aluminum [\onlinecite{N0}]), $V$ is the volume of the film, $k_\mathrm{B}$ is the Boltzmann constant, $T_c$ is the superconducting transition temperature (e.g., 1.2$\,$K for the aluminum), and $\Delta$ is the gap energy,
\begin{equation}    
2\Delta = 3.52 k_\mathrm{B} T_c.
\end{equation}
We change $T$ with several steps. In each step after temperature stabilization, we repeats the measurement of the resonant peak ($f_r$) through the transmittance spectrum as a function of the microwave frequency.  The difference of the resonant peak ($\delta f_r$) from the original position ($f_{r0}$) is converted into the phase difference ($\delta\theta$) based on the following formula: 
\begin{equation}
\label{eq:delta_theta}
\delta\theta = \frac{\delta\theta}{\delta f_r}\delta f_r = - \frac{4Q_{r0}}{f_{r0}}\delta{f_r}, 
\end{equation}
where $Q_{r0}$ is the quality factor of the resonance at the minimum temperature. We obtain $\mathrm{d}\theta/\mathrm{d}N_\mathrm{qp}$ through the measured parameters: $Q_{r0}, f_{r0}$, and $\delta f_r$ for each temperature.

However, this method has some issues for consideration. First, the variation of $T$ is too large compared with the conditions in real-world operations. It is typically $10\,\mathrm{mK}\sim100\,\mathrm{mK}$ variation. Second, it is difficult to estimate the uncertainty of $T$ because we cannot directly probe the MKID temperature. Third, stray lights due to the emissions in the cryostat, e.g., the heater, and a readout microwave power also create bias toward $N_\mathrm{qp}$ [\onlinecite{straylight, readout1, readout2, pieter}]. They generate a large bias for the responsivity measurement, at least for a some factors. Fourth, disadvantage of this method is that it takes a long time to change the temperature and stabilize the system.

A power spectral density (PSD) contains the responisivity in its formula [\onlinecite{pieter}],
\begin{equation}
\label{eq:PSD}
S_\theta (f) = \frac{4N_\mathrm{qp}\tau_\mathrm{qp}}{(1+(2\pi f\tau_\mathrm{qp})^2)(1+(2\pi f\tau_\mathrm{res})^2)}\left(\frac{\mathrm{d}\theta}{\mathrm{d}N_\mathrm{qp}}\right)^2 + X_\mathrm{system},
\end{equation}
where $S_\theta$ is the phase response, $f$ is the frequency of the detector response, $\tau_\mathrm{res}$ is the resonator ring time given by $\tau_\mathrm{res} = Q_r/\pi f_r$, and $X_\mathrm{system}$ is the noise of the readout system. As described in [\onlinecite{pieter}], the fitting for the measured PSD is in the way to obtain the parameters $\mathrm{d}\theta/\mathrm{d}N_\mathrm{qp}$, $\tau_\mathrm{qp}$ and $X_\mathrm{system}$.  The $Q_r$ and $f_r$ should be measured in advance through the resonant spectrum as a function of the microwave frequency.

The biggest concern with this method is the effects of a two-level system (TLS). The TLS creates a frequency-dependent noise structure, i.e., 1/f noise, which degrades the accuracy of fitting parameters unless we suppress it. Another consideration is the need to construct a low-noise setup for the suppression of $X_\mathrm{system}$.

It is known that $N_\mathrm{qp}$ depends on the power of readout microwaves [\onlinecite{readout1, readout2, pieter}]. This is because a tiny loss of microwaves locally warms up the resonator device. Based on this knowledge, we propose a method for the responsivity calibration. Figure $\ref{fig:response}$ illustrates the response of $N_\mathrm{qp}$ and $\theta$ to change in readout power from high to lower power at time $t = t_0$. The $N_\mathrm{qp}$ decreases with this change in the time scale of the detector response. The phase response as a function of time is formulated as follows:
\begin{figure}[t]
  \centering
  \includegraphics[width=8cm]{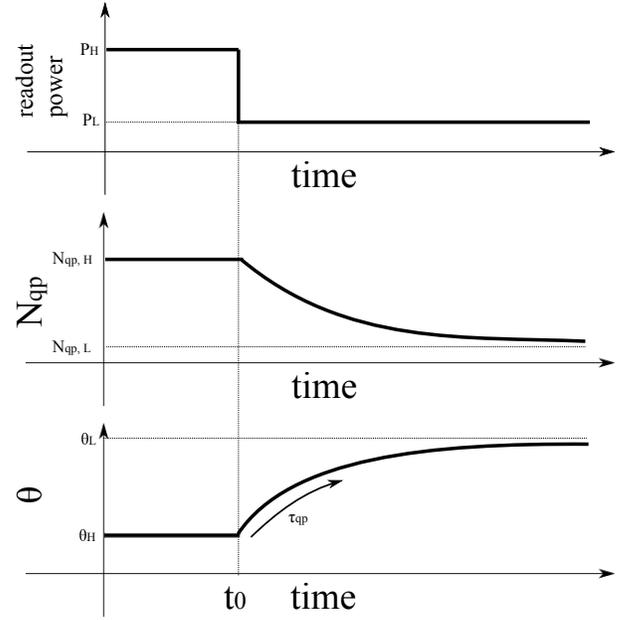}
  \caption{Illustrations of the principle to measure the responsivity, $\mathrm{d}\theta/\mathrm{d}N_\mathrm{qp}$, with changing power of the readout microwaves. The power change from $P_\mathrm{H}$ to $P_\mathrm{L}$ as illustrated in the top panel, causes the change of the $N_\mathrm{qp}$ with the time constant of the detector response, as illustrated in the middle. We observe this change in the phase response as illustrated at the bottom. }
  \label{fig:response}
\end{figure}
\begin{equation}
\label{eq:fit_theta}
\theta(t) = \begin{cases}
\theta_\mathrm{H} & (t < t_0) \\
 (\theta_\mathrm{H} - \theta_\mathrm{L})\exp\left(- \frac{t - t_0}{\tau_\mathrm{qp}}\right) + \theta_\mathrm{L} & (t \ge t_0),
 \end{cases}  
\end{equation}
where $\theta_\mathrm{H}$ and $\theta_\mathrm{L}$ are the phases before and after stabilizing the power change, respectively. This method allows us to measure the response, i.e., $\theta_\mathrm{L} - \theta_\mathrm{H}$, and $\tau_\mathrm{qp}$. As described in [\onlinecite{tauqp_Nqp}, \onlinecite{pieter2}], we obtain the $N_\mathrm{qp}$ using the following formula:
\begin{equation}
\label{eq:Nqp_tau_qp}
N_\mathrm{qp} = \frac{\tau_0 V}{\tau_\mathrm{qp}}\frac{N_0(k_\mathrm{B}T_c)^3}{2\Delta^2},
\end{equation}
where $\tau_0$ is the interaction time between electrons and photons (approximately 450$\,$ns for the aluminum [\onlinecite{tauqp_Nqp}, \onlinecite{pieter2}]). By using various sets of readout power changes, we obtain the phase response as a function of $N_\mathrm{qp}$, i.e., the responsivity. In this paper, we demonstrate this methodology by using hybrid-type MKIDs with aluminum film and niobium film. Note that this methodology should work in case of different types (e.g., multi-layer processing [\onlinecite{vissers}]) unless the quasiparticle lifetime is longer than several $\mu\mathrm{sec}$.

\begin{figure}[!b]
  \centering
  \includegraphics[width=8cm]{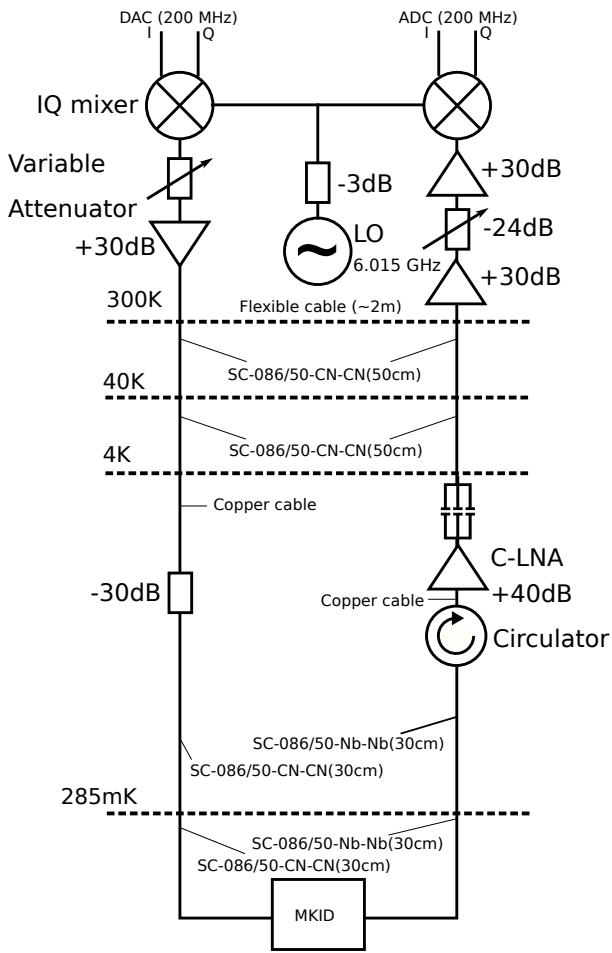}
  \caption{A diagram of the MKID readout. Our system generates feed signals at 200$\,$MHz bandwidth, which are up-converted and mixed in the microwave range. The local oscillator (LO) is the NI Microwave Components FSL-0010, and the details of the mixer are described in [\onlinecite{rfboard}]. We control the microwave power fed into the MKID using a variable attenuator as described in the text. Output microwaves from the MKIDs are amplified by the C-LNA (LNF-LN4$\_$8C s$/$n 176A, LOWNOISE FACTORY) and warm amplifiers (ZVE-8G+, Mini-Cirrcuits). After down-conversion of microwaves, we sampled them at 200$\,$MHz.}
  \label{fig:cryostat}
\end{figure}

We applied the proposed method in real measurements. The diagram of our setup is shown in Figure $\ref{fig:cryostat}$. Our cryostat consists of 4$\,$K, 40$\,$K, and 300$\,$K thermal shields [\onlinecite{GB_nagasaki}], and is cooled using a pulse tube refrigerator (PT415, Cryomech Co. Ltd). Magnetic shields (MS-FR, Hitachi material) were set outside of the 40$\,$K and 300$\,$K thermal shields; three sheets were set outside of the 40$\,$K shield, three sheets (four sheets) were set outside the wall (bottom plate) of the 300$\,$K shield. The MKID device was set in a light-tight box made of aluminum to mitigate the effects of stray light. The average temperature at the mounting plate of MKID was 285$\,$mK and was cooled by a helium-sorption refrigerator (Gas-Light type, Simon chase Co. LTD).  

Our MKID device was fabricated at RIKEN for astrophysical applications [\onlinecite{GB_nagasaki}]. This device consists of 10 hybrid-type MKIDs. The region for the input signal deposition is made of aluminum film, and other circuits are made of niobium film. The volume of the aluminum film is 920$\,\mu\mathrm{m}^3$ (the width is 4$\,\mu$m, the length is 2300$\,\mu$m, and the thickness is 100$\,$nm). The MKID used in this demonstration is not coupled to any antenna. Under a dark condition, its resonant frequency and quality factor are $f_{r0} = 6.07\,$GHz and $Q_{r0} = 4.78\times10^4$, respectively. Our readout electronics measured the resonance based on a direct down-conversion logic with 200$\,$MHz sampling speed [\onlinecite{RHEA_ishitsuka}, \onlinecite{RHEA_suzuki}], and the data was down-sampled to 1$\,$MHz step. This step is sufficient for this demonstration.

The power of the readout microwave was controlled using a variable attenuator (LDA-602E, Vaunix Co. LTD). It takes several $\,\mu$s to change the power after we set the attenuation value. For the responsivity calibration, we used five attenuation sets to change the microwave power from $P_\mathrm{H}$ to $P_\mathrm{L}$:  -11.0$\,$dB $\rightarrow$ -17.5$\,$dB, -12.0$\,$dB $\rightarrow$ -17.5$\,$dB, -13.0$\,$dB $\rightarrow$ -17.5$\,$dB,  -14.0$\,$dB $\rightarrow$ -17.5$\,$dB, and -15.0$\,$dB $\rightarrow$ -17.5$\,$dB. The feed power into the MKID is approximately $-70\,\mathrm{dBm}$ in the condition of $P_\mathrm{L}$.

The measured phase response ($\theta_\mathrm{measured}$) is the sum of the MKID signal ($\theta_\mathrm{MKID}$) and phase offset due to cables in the system ($\theta_\mathrm{system}$).
\begin{equation}
\label{eq:delay}
\theta_\mathrm{measured} = \theta_\mathrm{MKID} + \theta_\mathrm{system}.
\end{equation}
The $\theta_\mathrm{system}$ is a constant value unless the attenuation is changed. We measured it in the transmittance spectrum as a function of the microwave frequency for each power state, $P_\mathrm{H}$ and $P_\mathrm{L}$. In this responsivity calibration, we subtracted them in the analysis. We conservatively estimated the systematic error as 0.3$\,$rad. We also corrected the nonlinear effect of the response using the formula described in Appendix E of [\onlinecite{gao}],
\begin{equation}
\label{eq:correct}
\theta = 2\tan(\theta_\mathrm{MKID}/2).
\end{equation}

\begin{figure}[!b]
  \centering
  \includegraphics[width=9cm]{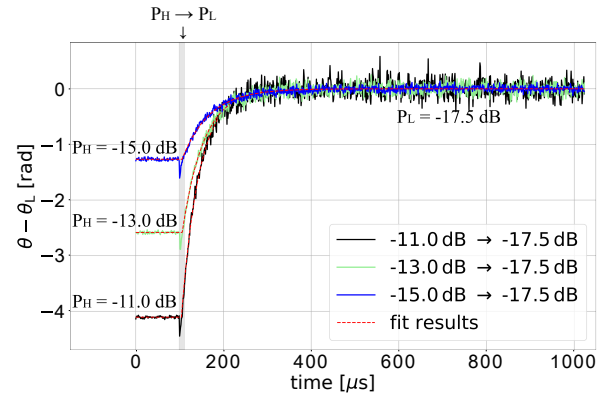}
  \caption{Measured phase responses as a function of time when we change the power of readout microwaves. It takes several $\mu$s to change the power after we reset the attenuation value at $t = 100\,\mu$s. Therefore, we do not use $t = 100\sim110\,\mu$s. We float $t_0$ as well as the phase difference and the quasiparticle lifetime when fitting this data. The dashed lines are the fitting results.}
    \label{fig:one_example}
\end{figure}
{\tabcolsep = 6mm
\begin{table*}[!t]
  \caption{Measured results for each readout power change. Errors are dominated by systematic errors described in the text. Statistical errors are smaller than systematic errors by approximately one order of magnitude.}\begin{tabular}{lccc} \hline
    $P_\mathrm{H}\rightarrow P_\mathrm{L}$ [dB]& $\theta_\mathrm{L} - \theta_\mathrm{H}\,$[rad] & $\tau_\mathrm{qp}\,[\mu$s] & $N_\mathrm{qp} /10^6$ \\ \hline
    -11.0 $\rightarrow$ -17.5 & $4.17 \pm 0.31$  & $41.6 \pm 1.0$   & $2.94 \pm 0.07$ \\
    -12.0 $\rightarrow$ -17.5 &$3.37 \pm 0.30$ & $43.4 \pm 1.0$  &  $2.82 \pm 0.06$ \\
    -13.0 $\rightarrow$ -17.5  &$2.60 \pm 0.30$ &$48.1 \pm 1.0$  &  $2.54 \pm 0.05$ \\ 
    -14.0 $\rightarrow$ -17.5  &$2.00 \pm 0.30$ &$54.9 \pm 1.0$  & $2.22 \pm 0.04$ \\
    -15.0 $\rightarrow$ -17.5  &$1.26 \pm 0.30$   &$60.7 \pm 1.0$ & $2.02 \pm 0.03$ \\ \hline
  \end{tabular}
  \label{tab:result}
\end{table*}
}
Figure $\ref{fig:one_example}$ shows phase responses as a function of time. We reset the attenuation value at $t = 100\,\mu$s with a precision of 1$\,\mu$s. We observe the response in the phase. We fit the data to Eq. ($\ref{eq:fit_theta}$), and extract parameters. We did not use the data in the short period $t = 100\,\mu\mathrm{s}\sim110\,\mu\mathrm{s}$ because of the uncontrolled state of attenuation after the reset, as mentioned above. We repeat this measurement 40$\,$ times for each set of power change. The results are summarized in Table $\ref{tab:result}$. 

\begin{figure}[!t]
  \centering
  \includegraphics[width=9cm]{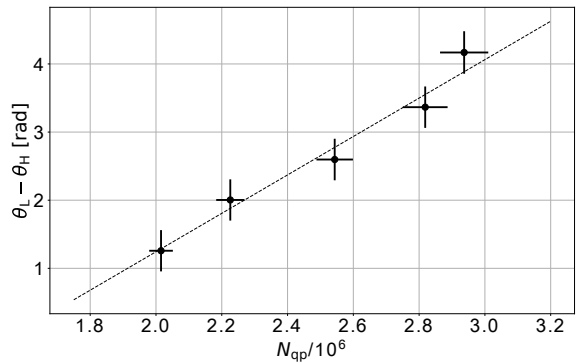}
  \caption{The measured relation between the phase response ($\theta_\mathrm{L} - \theta_\mathrm{H}$) and the number of quasiparticles ($N_\mathrm{qp}$) based on the proposed method. We obtain the responsivity from the linear fit.}
  \label{fig:result}
\end{figure}

Figure $\ref{fig:result}$ shows the relation between the phase response as a function of $N_\mathrm{qp}$. This is based on values listed in Table $\ref{tab:result}$. We obtain $\mathrm{d}\theta/\mathrm{d}N_\mathrm{qp} = (2.8 \pm 0.3) \times 10^{-6}\,$rad from this plot, assuming the linear relation.

For comparison, we also perform conventional methods: the heater control method and PSD method. With the heater control method, we changed the temperature of the mounting plate for the MKID from 285$\,$mK to 300$\,$mK, and obtained $\mathrm{d}\theta/\mathrm{d}N_\mathrm{qp} = (9.9\pm0.3)\times10^{-7}\,$rad, here we only assign the statistical error. The systematic uncertainty of this result is a few factors, as mentioned above. With the PSD method, we obtained $\mathrm{d}\theta/\mathrm{d}N_\mathrm{qp}=(2.4\pm0.2)\times10^{-6}\,$rad. The bias due to the 1/f noise was not evaluated and it was not included. However, the 1/f noise tends to generate small bias. Comparing to systematic uncertainties in each method, discrepancies among the three results are not significant.

In summary, MKID is one of  cutting-edge superconducting detector based on the principle of superconducting resonator in the microwave range. The responsivity calibration, i.e., understanding the relation between the response in the resonant phase and number of quasiparticles, is important for both its application and development. We proposed a calibration method, i.e., simultaneous  measurements of the phase difference and the quasiparticle lifetime following the power change of the readout microwave. The number of quasiparticles was calculated using the measured lifetimes and geometrical parameters of the detector device. We demonstrated this method, and confirmed its simplicity and ease-of-use. We also performed conventional methods: the heater control method and PSD method. We confirmed the consistency between the results obtained using our method and conventional methods in terms of accuracy. Therefore, our method is promising for the responsivlity calibration. 
\\

This work was supported by the JRA program in RIKEN, and Grants-in-Aid for Scientific Research from The Ministry of Education, Culture, Sports, Science and Technology, Japan (KAKENHI Grants. 15H05743, 16J09435, 18H05539, 15K13491, 16H00874, and R2804). We thank  Shunsuke Honda, Taketo Nagasaki, and Junya Suzuki for useful discussions. We thank Mr. Noboru Furukawa, Advanced Technology Center of National Astronomical Observatory of Japan, and Advanced Manufacturing Support Team of RIKEN.

\nocite{*}
\bibliography{AIP_dtheta_dP_paper}

\providecommand{\noopsort}[1]{}\providecommand{\singleletter}[1]{#1}%
\begin{thebibliography}{10}

\bibitem{MKIDs}
{P. K. Day, H. G. Leduc, B. A. Mazin, A. Vayonakis, and J. Zmuidzinas}.
\newblock {\em Nature}, 425(6960):817, 2003.

\bibitem{DESHIMA}
{A. Endo, P. Werf, R. M. J. Janssen, P. J. Visser, T. M. Klapwijk, J. J. A.
  Baselmans, L. Ferrari, A. M. Baryshev, and S. J. C. Yates}.
\newblock {\em Journal of Low Temperature Physics}, 167(3-4):341--346, 2012.

\bibitem{NIKA}
{A. Monfardini, L. J. Swenson, A. Bideaud, F. X. D\'{e}sert, S. J. C. Yates, A.
  Benoit, A. M. Baryshev, J. J. A. Baselmans, S. Doyle, B. Klein, M. Roesch, C.
  Tucker, P. Ade, M. Calvo, P. Camus, C. Giordano, R. Guesten, C. Hoffmann, S.
  Leclercq, P. Mauskopf, and K. F. Schuster}.
\newblock {\em Astronomy \& Astrophysics}, 521:A29, 2010.

\bibitem{GB_nagasaki}
{T. Nagasaki, J. Choi, R. T. G\'{e}nova-Santos, M. Hattori, H. Hazumi, K.
  Ishitsuka, K. Karatsu, K. Kikuchi, R. Koyano, H. Kutsuma, K. Lee, S. Mima, M.
  Minowa, M. Nagai, M. Naruse, S. Oguri, C. Otani, R. Rebolo, J. A.
  Rubi\~{n}o-Mart\'{i}n, Y. Sekimoto, M. Semoto, J. Suzuki, T. Taino, O.
  Tajima, N. Tomita, T. Uchida, E. Won, and M. Yoshida}.
\newblock {\em Journal of Low Temperature Physics}, 193(5-6):1066--1074, 2018.

\bibitem{tauqp_Nqp}
{S. B. Kaplan, C. C. Chi, D. N. Langenberg, J. J. Chang, S. Jafarey, and D. J.
  Scalapino}.
\newblock {\em Physical Review B}, 14(11):4854, 1976.

\bibitem{pieter2}
{P. J. de Visser, J. J. A. Baselmans, P. Diener, S. J. C. Yates, A. Endo, and
  T. M. Klapwijk}.
\newblock {\em Physical review letters}, 106(16):167004, 2011.

\bibitem{gao2008}
{J. Gao, J. Zmuidzinas, A. Vayonakis, P. Day, B. Mazin, and H. Leduc}.
\newblock {\em Journal of Low Temperature Physics}, 151(1-2):557--563, 2008.

\bibitem{BCS}
{J. Bardeen, L. N. Cooper, and J. R. Schrieffer}.
\newblock {\em Physical review}, 108(5):1175, 1957.

\bibitem{N0}
{W. L. McMillan}.
\newblock {\em Physical Review}, 167(2):331, 1968.

\bibitem{straylight}
J.~J.~A. Baselmans and S.~J.~C. Yates.
\newblock {\em AIP Conference Proceedings}, 1185(1):160--163, 2009.

\bibitem{readout1}
{S. E. Thompson, S. Withington, D. J. Goldie, and C. N.Thomas}.
\newblock {\em Superconductor Science and Technology}, 26(9):095009, 2013.

\bibitem{readout2}
{P. J. de Visser, D. J. Goldie, P. Diener, S. Withington, J. J. A.Baselmans,
  and T. M. Klapwijk}.
\newblock {\em Physical review letters}, 112(4):047004, 2014.

\bibitem{pieter}
{P. J. de Visser}.
\newblock {\em Quasiparticle dynamics in aluminium superconducting microwave
  resonators}.
\newblock PhD thesis, Delft University of Technology, Delft, The Netherlands,
  2014.

\bibitem{vissers}
{M. R. Vissers, J. Gao, M. Sandberg, S. M. Duff, D. S. Wisbey, K. D.Irwin, and
  D. P. Pappas}.
\newblock {\em Applied Physics Letters}, 102(23):232603, 2013.

\bibitem{rfboard}
{J. v. Rantwijk, M. Grim, D. v. Loon, S. Yates, A. Baryshev, and J. Baselmans}.
\newblock {\em IEEE Transactions on Microwave Theory and Techniques},
  64(6):1876--1883, 2016.

\bibitem{RHEA_ishitsuka}
{H. Ishitsuka, M. Ikeno, S. Oguri, O. Tajima, N. Tomita, and T. Uchida}.
\newblock {\em Journal of Low Temperature Physics}, 184(1-2):424--430, 2016.

\bibitem{RHEA_suzuki}
{J. Suzuki, H. Ishitsuka, K. Lee, S. Oguri, O. Tajima, N. Tomita, and E. Won}.
\newblock {\em Journal of Low Temperature Physics}, 193(3-4):562--569, 2018.

\bibitem{gao}
{J. Gao}.
\newblock {\em The physics of superconducting microwave resonators}.
\newblock PhD thesis, California Institute of Technology, 2008.

\end{thebibliography}
\bibliographystyle{unsrt}

\end{document}